%% file: hexagonal_bic_symmetry.tex
\documentclass[%
 reprint,
 amsmath,amssymb,
 aps,
prb,
]{revtex4-2}

\usepackage{graphicx}% Include figure files
\usepackage{dcolumn}% Align table columns on decimal point
\usepackage{bm}% bold math
\usepackage{caption}
\usepackage{subcaption}
\usepackage{glossaries}
\usepackage{braket}
\usepackage{siunitx}
\input{glossary_entries}
\usepackage[hidelinks]{hyperref}% add hypertext capabilities
\DeclareMathOperator{\Real}{Re}
\DeclareMathOperator{\Imag}{Im}

\makeatletter
\expandafter\let\csname longtable*\endcsname\relax
\expandafter\let\csname endlongtable*\endcsname\relax
\AddToHook{begindocument/before}{%
    \expandafter\let\csname longtable*\endcsname\relax
    \expandafter\let\csname endlongtable*\endcsname\relax
}
\makeatother

\begin{document}

\preprint{APS/123-QED}

\title{\textbf{Symmetry Properties of Bound States in the Continuum in Hexagonal Lattice Photonic Crystal Slabs} 
}% 

\author{Rhys Jones}
\author{Sang Soon Oh}%
 \email{Contact author: OhS2@cardiff.ac.uk}
\affiliation{%
School of Physical, Chemical, and Environmental Sciences, Cardiff University, CF24 3AA, United Kingdom
}%

\date{\today}% It is always \today, today,

\begin{abstract}
\Glspl{bic} within \gls{phc} cavities realise the full confinement of light despite lying within the continuum of radiation modes. While their diverse practical applications has attracted considerable research interest, developing the fundamental understanding of \glspl{bic} is important for the design of future devices. This paper uses a group theoretical approach to determine the basis functions of the \glspl{irrep} of a \gls{phc} with the symmetry of the $C_{6v}$ space group, from which the spatial distribution of the electromagnetic fields within the cavity can be determined. Furthermore, the reflection parity of the electric field and a $\bm{k} \cdot \bm{p}$ perturbation theory approach are used to determine the far-field polarisation of each \gls{irrep}. The $\bm{k} \cdot \bm{p}$ perturbation explains why the $E_2$ \gls{irrep} has a topological charge of $-2$, as well as showing that a topological charge does not necessarily indicate the presence of a \gls{bic}. Additionally, it is found that the quality factor of the modes do not necessarily scale as $Q \propto 1/\left|k\right|^{2\left|q\right|}$, but are instead dependent on the coupling to the radiative mode in the $\bm{k} \cdot \bm{p}$ perturbation, which is determined from the symmetry of each \gls{irrep}. 
\end{abstract}

\maketitle

%\tableofcontents

\glsresetall
\section{\label{sec:introduction}Introduction}
\Glspl{bic} are states that are localised within a cavity, despite their frequency lying within the continuum of radiation modes \cite{hsu2016}. Due to their perfectly bound nature, \glspl{bic} have infinite \gls{q} factor, which has lead to diverse interest in their applications for low threshold lasing \cite{kodigala2017}, semiconductor manufacturing \cite{zhang2026}, and biological sensing \cite{kang2023}. For further development and application of practical devices, it is important to develop the fundamental understanding of the nature of \glspl{bic}.

\begin{figure*}[ht]
    \centering
    \hfill
    \begin{subfigure}[b]{0.2\linewidth}
        \begin{minipage}[b][4.1cm][c]{\linewidth}
            \centering
            \includegraphics[width=\linewidth]{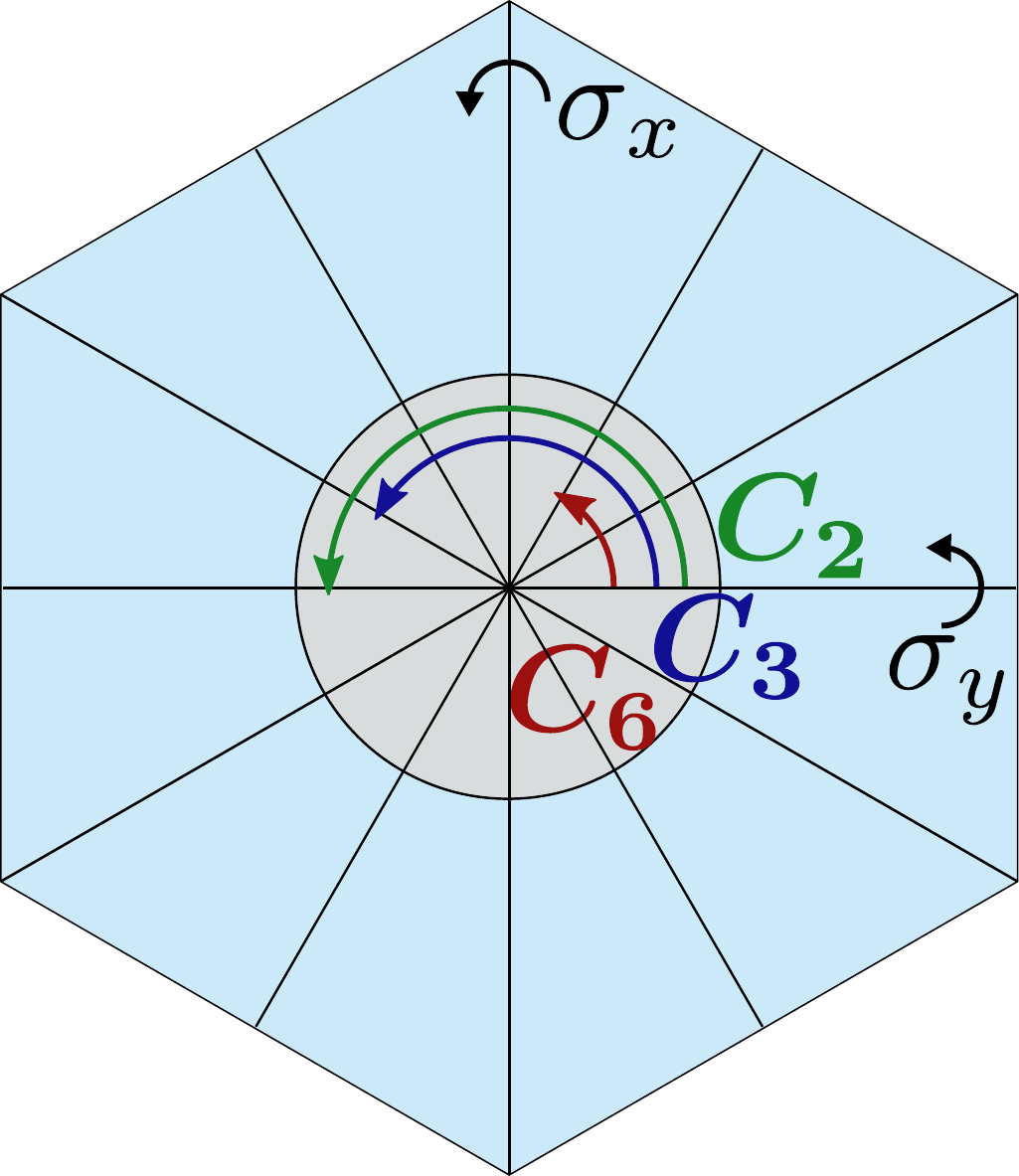}
        \end{minipage}    
        \caption{}
        \label{fig:real_symmetries}
    \end{subfigure}
    \hfill
    \begin{subfigure}[b]{0.25\linewidth}
        \begin{minipage}[b][4.1cm][c]{\linewidth}
            \centering
            \includegraphics[width=\linewidth]{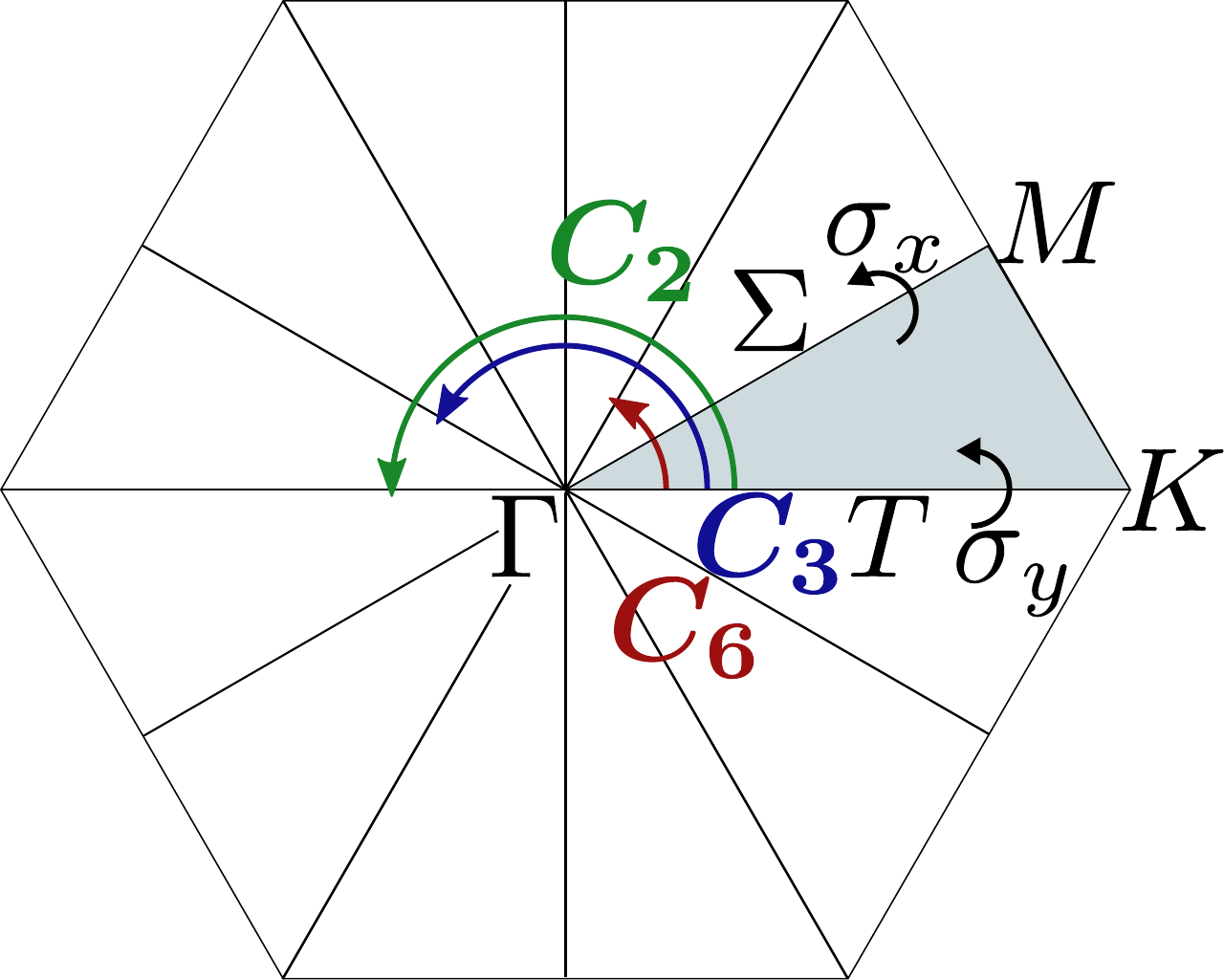}
        \end{minipage}
        \caption{}
        \label{fig:reciprocal_symmetries}
    \end{subfigure}
    \hfill
    \begin{subfigure}[b]{0.42\linewidth}
        \begin{minipage}[b][4.1cm][c]{\linewidth}
            \centering
            \includegraphics[width=\linewidth]{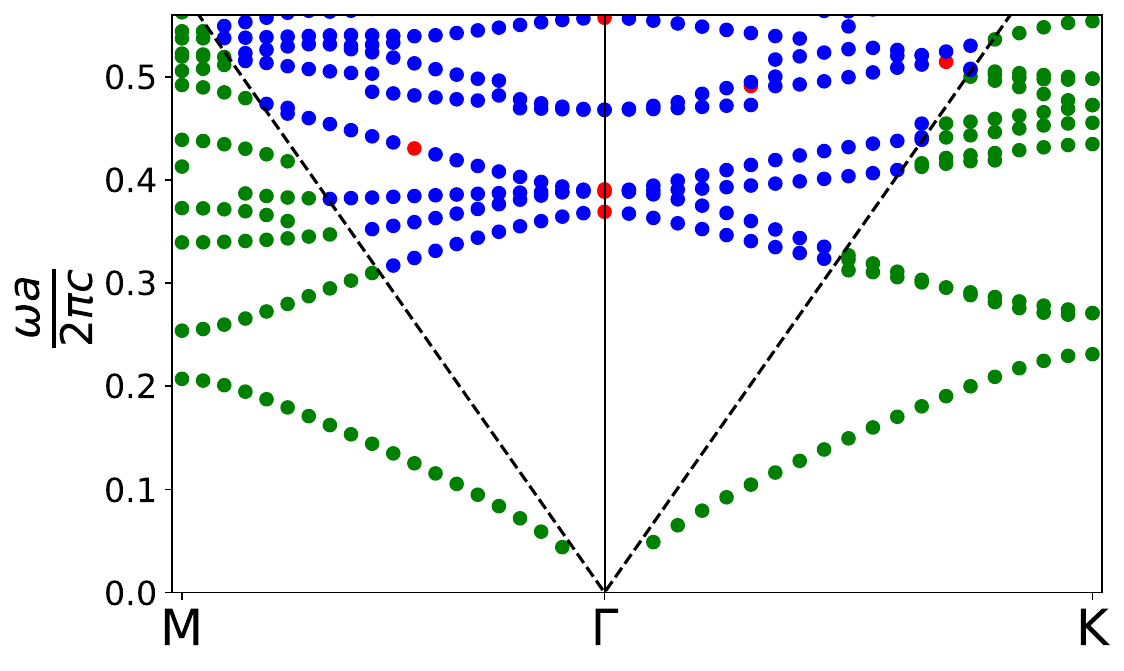}
        \end{minipage}
        \caption{}
        \label{fig:band_structure}
    \end{subfigure}
    \hfill
    \vspace{1em}

    \hfill
    \begin{subfigure}[b]{0.23\linewidth}
        \begin{minipage}[b][3.6cm][c]{\linewidth}
            \centering
            \includegraphics[width=\linewidth]{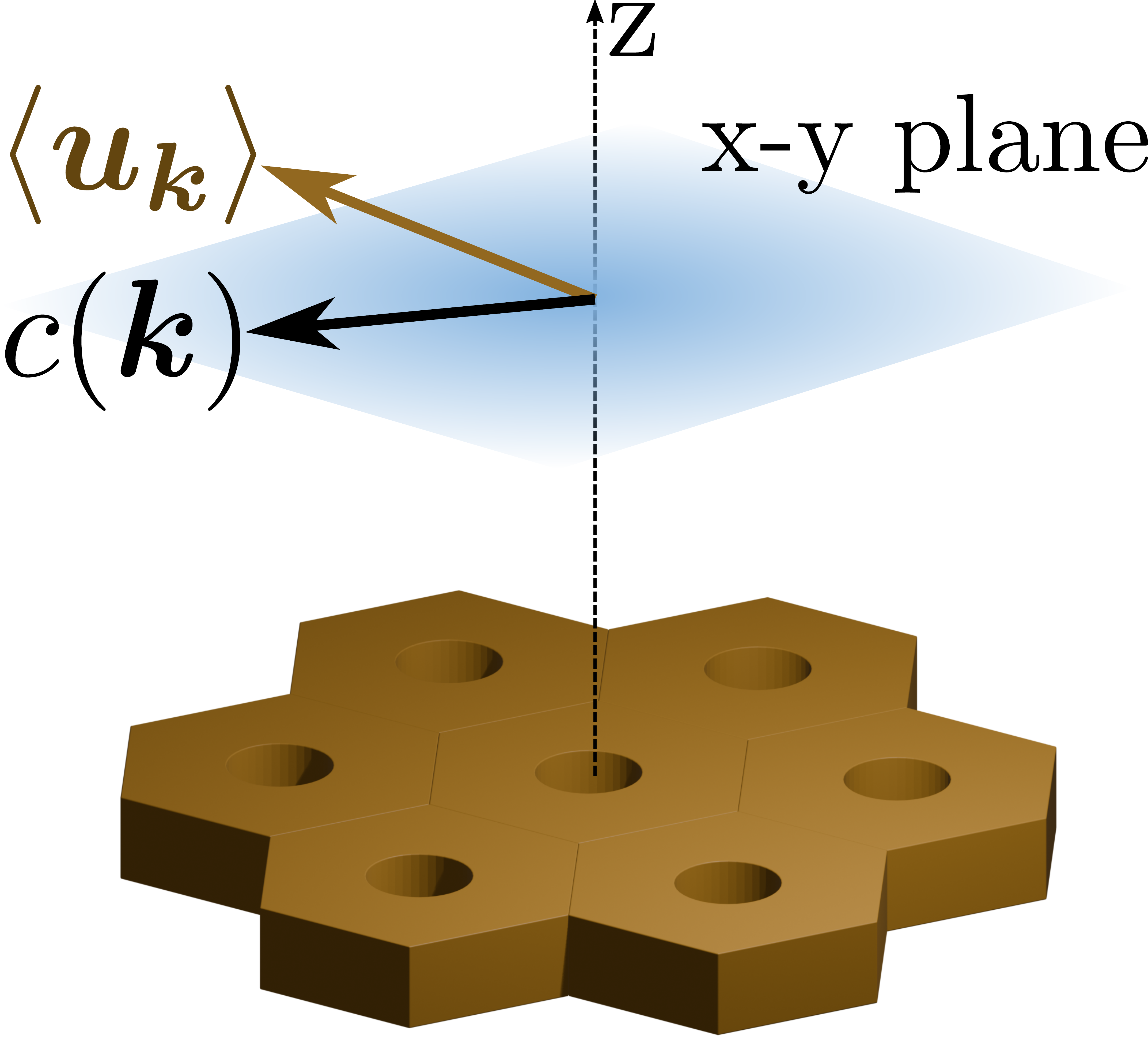}
        \end{minipage}
        \caption{}
        \label{fig:far_field_explanation}
    \end{subfigure}
    \hfill
    \begin{subfigure}[b]{0.7\linewidth}
        \begin{minipage}[b][3.6cm][c]{\linewidth}
            \centering
            \includegraphics[width=\linewidth]{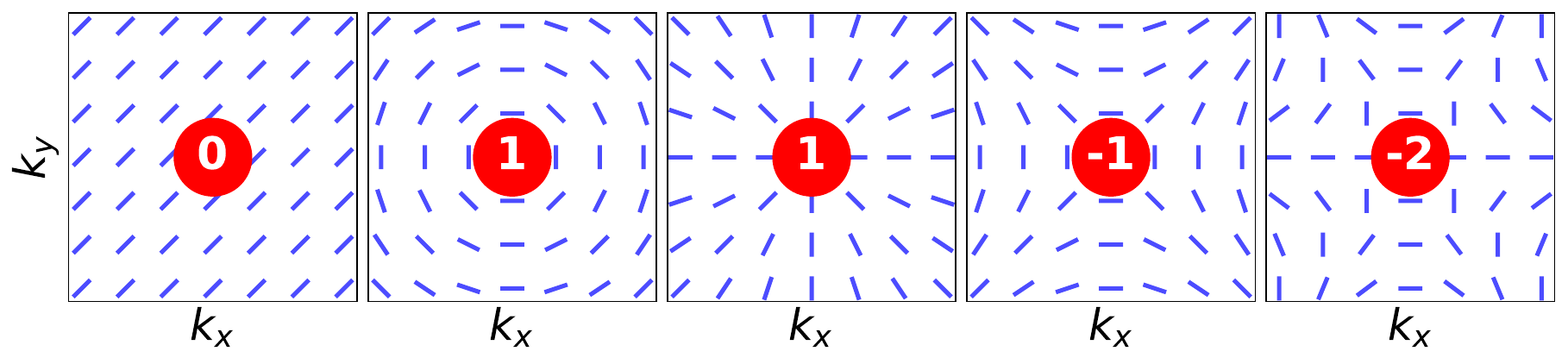}
        \end{minipage}
        \caption{}
        \label{fig:tc_explanation}
    \end{subfigure}
    \hfill
    \hspace{0.1em}
    \caption{(\ref{sub@fig:real_symmetries}) and (\ref{sub@fig:reciprocal_symmetries}) The symmetries of a \glsentryshort{phc} belonging to the $C_{6v}$ group in real and reciprocal space respectively. High symmetry points and lines in reciprocal space are shown. (\ref{sub@fig:band_structure}) Band structure of the same \glsentryshort{phc} calculated using the \gls{fem} method in COMSOL Multiphysics. Simulation details are provided in Section \ref{sec:numerical_simulations}. (\ref{sub@fig:far_field_explanation}) Illustrative diagram explaining the method for calculating the far-field polarisation. (\ref{sub@fig:tc_explanation}) Far-field polarisation distributions for different \glspl{tc}.}
    \label{fig:illustrative}
    \glsreset{tc} \glsreset{fem}
\end{figure*}

\Glspl{bic} can be split into two main categories, depending upon the method of their formation. Symmetry-protected \glspl{bic} are formed when the symmetry of the bound mode in the cavity is different to the symmetry of the far-field radiation \cite{hsu2016}, which results in the bound mode being unable to couple to the far-field radiation. They are most commonly found at the $\Gamma$ point, which conforms to the full rotational symmetry of the structure \cite{sakoda2005}. For a structure with $C_{6v}$ symmetry and below the first diffraction order \cite{wang2026a} the diffracted waves have two polarisation components which transform as the $\hat{x}$ and $\hat{y}$ unit vectors, conforming to the symmetry of the $E_1$ \gls{irrep} \cite{ochiai2001}. Therefore, while the $E_1$ \gls{irrep} can couple to the far-field radiation, all other \glspl{irrep} of the space group are prohibited from coupling to the far-field due to mismatched symmetry and become symmetry-protected \glspl{bic}. 

The second type of \glspl{bic} are accidental-\glspl{bic}, which generally occur at off-$\Gamma$ points. They are formed due to destructive interference between two constituent modes, where the lifetime of one mode is reduced, while the radiation of the second mode is subject to total destructive interference leading to a state with an infinite lifetime, known as a \gls{bic} \cite{hsu2016}. Accidental \glspl{bic} can be considered as Friedrich-Wintgen \glspl{bic} and can be formed from either the destructive interference of two guided resonance modes \cite{lee2020}, or from the interference between a guided resonance and a Fabry–Pérot mode \cite{hu2022}. The different types of resonances supported by a \gls{phc} cavity is shown in the band structure in Figure \ref{fig:band_structure}. States which occur below the light line and are unable to couple to the far-field are shown in green, while states that are able to couple to the far-field are shown in blue. The high-\gls{q} \glspl{bic} are seen in red, where the states at the $\Gamma$ point are symmetry-protected \glspl{bic}, and the off-$\Gamma$ \glspl{bic} are the accidental-\glspl{bic}. 

The properties of symmetry-protected \glspl{bic} can be derived using a group theoretical approach. A \gls{phc} can be classified according to its rotational, reflection and translational symmetries, the set of which forms the space group, $\mathcal{M}$, of the structure. While the $\Gamma$ point conforms to the full symmetry of the structure, the reduced symmetry group at an arbitrary $\bm{k}$ vector is given by $\mathcal{M}_{\bm{k}}$. All members, $R$ of the group $\mathcal{M}_{\bm{k}}$ commute with the differential operator that solves Maxwell's equations for a \gls{te} polarisation \cite{sakoda2005}:
\begin{equation}
    R \mathcal{L}_H^{(2)} R^{-1} = \mathcal{L}_H^{(2)},
\end{equation}
therefore any eigenfunction of $\mathcal{L}_H$ is also an eigenfunction of $R$ \cite{sakoda1995}. We are also able to state that any eigenfunction of $\mathcal{L}_H^{(2)}$ is an \gls{irrep} of the group $\mathcal{M}_{\bm{k}}$. The symmetry of the \glspl{irrep} are tabulated in character tables. From the symmetries under group operations it is possible to determine the basis functions of each of the \glspl{irrep} \cite{dresselhaus2008}, which corresponds to the out of plane scalar field in a two-dimensional \gls{phc} structure. Using a perturbative approach for $\bm{k}$ vectors near the $\Gamma$ point \cite{mei2012} a group theoretical approach can also be used to determine the far-field polarisation near a \gls{bic} \cite{ochiai2024}.

\glspl{bic} can also be understood as polarisation vortices in momentum space. Their vortex nature is a consequence of the fact that \glspl{bic} are fully confined within a cavity and therefore have zero far-field polarisation. The polarisation vector winds around the polarisation vortex, and the winding of the far-field polarisation vector determines the \gls{tc}, $q$, of the structure \cite{zhen2014}. Illustrative diagrams of the far-field polarisation of \glspl{bic} are shown in Figure \ref{fig:far_field_explanation} and \ref{fig:tc_explanation}. While the treatment of the \gls{tc} of non-degenerate linearly polarised modes is straightforward and has been extensively covered in the literature \cite{hsu2016, zhen2014, kang2022, salerno2022}, the description of degenerate and circularly polarised states are more complex. Specifically, for a structure with $C_{6v}$ symmetry while experimental and numerical results have shown the \gls{tc} of the $E_2$ ($E_1$) mode to be -2 (1) \cite{ratiu2026, zhao2024, chern2026}, theoretical approaches have shown the charges to be -1 (0) \cite{salerno2022}. A \gls{tc} of 0 for the $E_1$ mode is also consistent with the traditionally expected value of the \gls{tc} for a non-\gls{bic} radiative mode. 

In this paper, the symmetry properties of each of the \glspl{irrep} of the $C_{6v}$ group will be introduced. First the near-field properties of each \gls{irrep} will be determined using basis functions from an angular momentum principle. Then the far-field polarisation will be determined using the reflection parity of the electric field along high symmetry lines and the $\bm{k} \cdot \bm{p}$ perturbation theory. Both methods will be used to determine the \gls{tc} of each mode. Meanwhile, by looking at the coupling to the radiative mode in the $\bm{k} \cdot \bm{p}$ perturbation theory, the scaling of the \gls{q} factor at $\bm{k}$ vectors near the $\Gamma$ point will also be determined. All results will then be verified using numerical simulation methods.

\begin{figure*}[ht]
    \centering
    \includegraphics[width=0.9\linewidth]{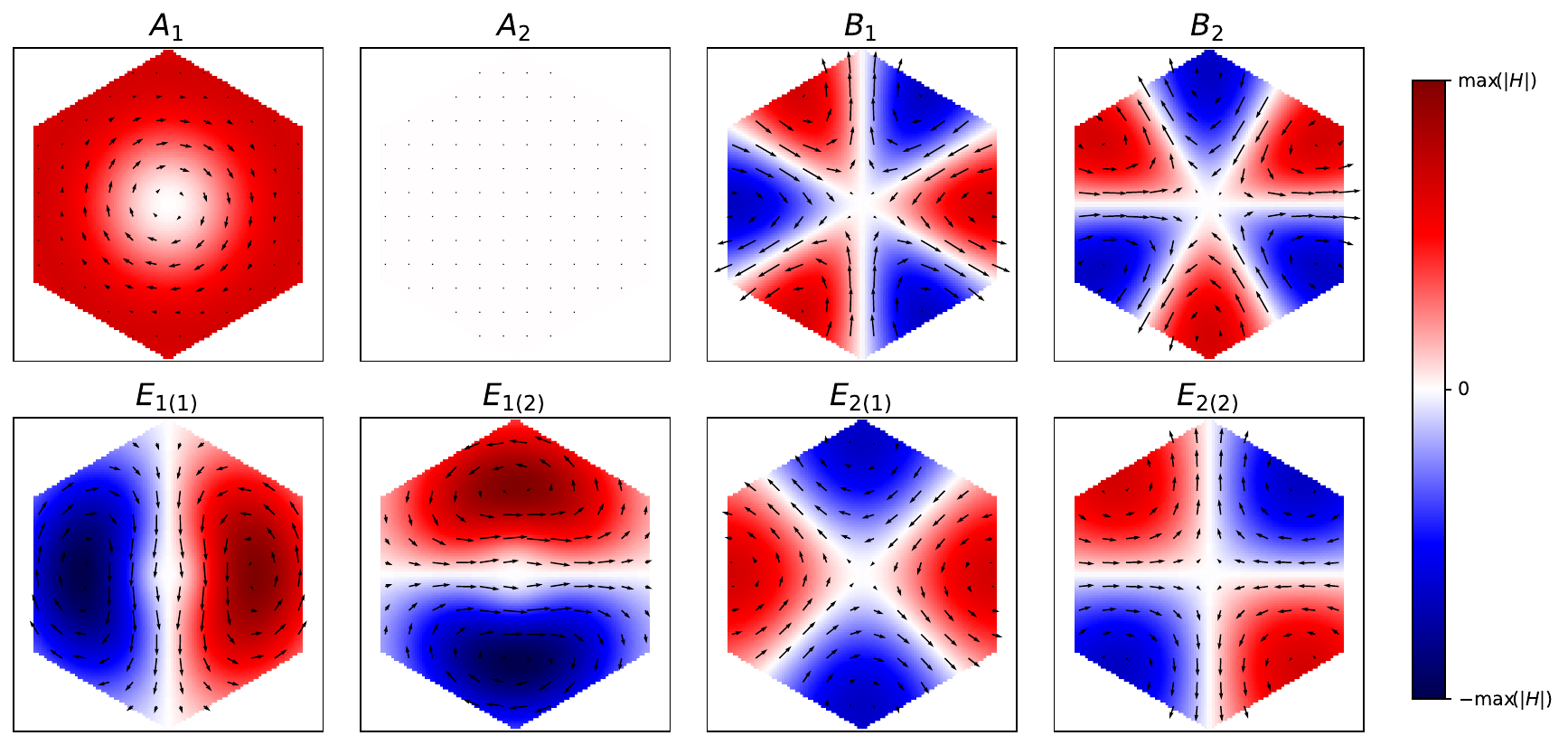}
    \caption{Electromagnetic fields within a unit cell for each of the 6 \glspl{irrep} with $n=0$. Colour plot shows the imaginary part of the magnetic field. Arrows represent the real part of the electric field. While any linear combination of $E_{i(1)}$ and $E_{i(2)}$ for $i=1, 2$ are an acceptable basis function, the angular momentum basis vectors $(x, y)$ and $(x^2-y^2, -2xy)$ have been chosen as the basis functions for convenience.}
    \label{fig:c6v_fields}
\end{figure*}

\section{Near-Field Properties}
A \gls{phc} with a hexagonal unit cell and a circular air hole in the middle has the symmetry of the $C_{6v}$ space group \cite{sakoda2005}. The structure is therefore unchanged under an identity operation, a $C_6, C_3$ or $C_2$ rotation, as well as under a reflection of the $x$ or $y$ coordinates, where all operations except for the identity are shown in Figure \ref{fig:real_symmetries}. While the focus of this paper will be on a hexagonal unit cell with a single air hole note that, provided each atom in the unit cell are identical, both the honeycomb and Kagome lattices belong to the same symmetry group and therefore have identical symmetry properties. In reciprocal space the $\Gamma$ point has the same symmetry as the full $C_{6v}$ space group, while the $K$ and $M$ points have $C_{3v}$ and $C_{2v}$ symmetries, and the $T$ and $\Sigma$ lines have the symmetry of the $C_{1h}$ point group \cite{sakoda2005}. The high symmetry points and lines are shown in Figure \ref{fig:reciprocal_symmetries}. 

The character table of the $C_{6v}$ space group is shown in Table \ref{tab:c6v_characters} \cite{atkins1970}, where the $A$ and $B$ modes are non-degenerate, while the $E_1$ and $E_2$ modes are both doubly degenerate.
\begin{table}[ht] 
    \renewcommand{\arraystretch}{1.2}
    \begin{tabular}{c|cccccc} 
        \hline $C_{6v}$ & $E$ & $2C_6$ & $2C_3$ & $C_2$ & $3\sigma_y$ & $3\sigma_x$ \\ 
        \hline $A_1$ & 1 & 1 & 1 & 1 & 1 & 1 \\ 
        $A_2$ & 1 & 1 & 1 & 1 & -1 & -1 \\ 
        $B_1$ & 1 & -1 & 1 & -1 & 1 & -1 \\ 
        $B_2$ & 1 & -1 & 1 & -1 & -1 & 1 \\ 
        $E_1$ & 2 & 1 & -1 & -2 & 0 & 0 \\ 
        $E_2$ & 2 & -1 & -1 & 2 & 0 & 0 \\ \hline 
    \end{tabular} 
    \caption{Character table of the point group $C_{6v}$.} 
    \label{tab:c6v_characters}
\end{table}
The table can be used to determine the symmetry of the out-of-plane scalar field under a symmetry operation of the group. In this paper the character table informs us about the nature of the magnetic field as we focus on the \gls{te} mode, but equivalent relations may be derived for transverse magnetic modes. The in-plane electric field transforms with the same character as the magnetic field under a rotation, but transforms in an inverted manner under reflection due to the coordinate transformation. 

An intuitive method for determining the form of the magnetic field for each of the \glspl{irrep} is by using an angular momentum basis \cite{wu2015}. Each of the basis vectors can be determined from $\exp[i(6n+m)\theta]$, where the lowest order modes occurring at $n=0$ will be considered here. Each unique value of $m$ corresponds to a different \gls{irrep}, with $m=0$ representing the $A$ modes, $m=1$ the $E_1$ mode, $m=2$ the $E_2$ mode, while $m=3$ represents the $B$ modes. For the non-degenerate modes the real component represents the mode that is symmetric under the primary reflection (represented by a subscript of 1 in Table \ref{tab:c6v_characters}), while the imaginary component represents the antisymmetric under primary reflection mode. For the degenerate modes any linear combination of the real and imaginary component is an acceptable eigenfunction of the mode. Figure \ref{fig:c6v_fields} plots the lowest order mode of each of the \glspl{irrep} (note that only higher order modes exist for the $A_2$ \gls{irrep}). For visualisation the basis function is multiplied with a Maxwell distribution which was used to represent the effect of the refractive index distribution on the modes. The exact distribution for any structure can be determined with numerical simulation methods, as seen in Section \ref{sec:numerical_simulations}. The electric field distribution is calculated as $\bm{E} = \frac{i}{\omega \epsilon_0} \nabla \times \bm{H}$ using the Ampère-Maxwell law. 

From looking at the $B_2$ mode in Figure \ref{fig:c6v_fields} we can see that the magnetic field is invariant under an identity operation, a $C_3$ rotation and an inversion of the $x$ coordinate. On the other hand, under a $C_6$ or $C_2$ rotation, or an inversion of the $y$ coordinate the magnetic field flips sign, conforming to the character table in Table \ref{tab:c6v_characters}. It can be verified that the other \glspl{irrep} in Figure \ref{fig:c6v_fields} also conform to Table \ref{tab:c6v_characters}. 

\section{Far-field Properties}
The far-field polarisation of an eigenmode is defined as $\bm{c(k)} = c_x(\bm{k}) \hat{x} + c_y(\bm{k}) \hat{y} = \hat{x} \cdot \left<\bm{u_k}\right> \hat{x} + \hat{y} \cdot \left<\bm{u}_k\right> \hat{y}$, shown in Figure \ref{fig:far_field_explanation}. The Stokes parameters of the far-field polarisation can be calculated as $S_0 = |c_x|^2 + |c_y|^2$, $S_1 = |c_x|^2 - |c_y|^2$, $S_2 = 2\mathrm{Re}(c_x c_y^*)$, and $S_3 = -2\mathrm{Im}(c_x c_y^*)$. Defining the angle of the far-field polarisation as $\psi(k_\parallel) = \arg(S_1 + iS_2)/2$, the angle may become undefined in three situations \cite{weimin2024}. Firstly the magnitude of the far-field polarisation can become 0, that is $S_0 = S_1 = S_2 = S_3 = 0$, which corresponds to a non-degenerate \gls{bic} or a V point in parameter space. Secondly we could have a circularly polarised far-field, represented by $S_1 = S_2 = 0$, $S_3 = \pm 1$, corresponding to a C point in parameter space. Lastly we could have an undefined $c_x c_y^*$, which corresponds to a degenerate state where any arbitrary combination of the basis vectors fulfil the eigenvalue condition \cite{long2023}; defined as a D point in parameter space. Defining a T point to be any of a V, C, or D point, we can calculate the \gls{tc}, $q$, by integrating in a counter-clockwise loop in reciprocal space around the T point \cite{zhen2014}:
\begin{equation}
    q = \dfrac{1}{2\pi} \oint_C \nabla_{k_\parallel} \psi(k_\parallel) \cdot dk_\parallel .
\end{equation}

All C points have a half-integer \gls{tc}, while the V and D point possess an integer \gls{tc}. While the \gls{tc} of the non-degenerate modes can be determined from the phase factor of their eigenvalue under a $C_6$ rotation \cite{salerno2022}, such a method cannot be used for the D points due to their undefined far-field polarisation. The \gls{tc} of the D points can therefore be determined in two alternative ways. The first approach uses the reflection parity of the far-field polarisation along the high symmetry $\Sigma$ and $T$ lines to determine the \gls{tc} of the structure, an approach that can also be used for the V points to verify its accuracy. The second method uses a $\bm{k} \cdot \bm{p}$ perturbation to determine both the \gls{tc} of the modes and the scaling of the \gls{q} factor as the $\bm{k}$ vector is moved away from the $\Gamma$ point. 

\subsection{Reflection Parity}
\begin{figure}[ht]
    \centering
    \begin{subfigure}{0.48\linewidth}
        \centering
        \includegraphics[width=\linewidth]{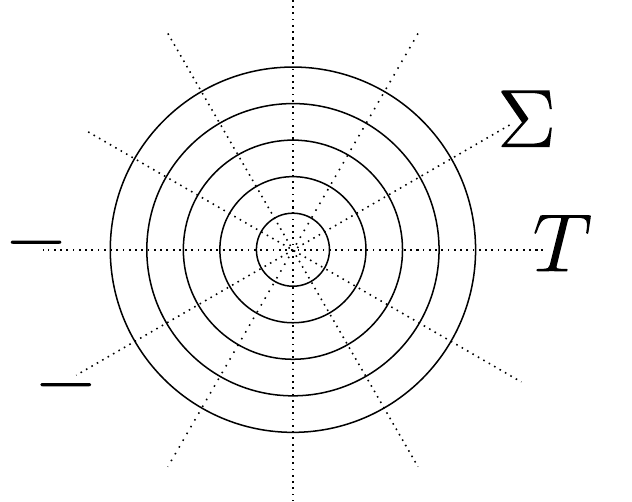}
        \caption{}
        \label{fig:parity_A1}    
    \end{subfigure}
    \hfill
    \begin{subfigure}{0.48\linewidth}
        \centering
        \includegraphics[width=\linewidth]{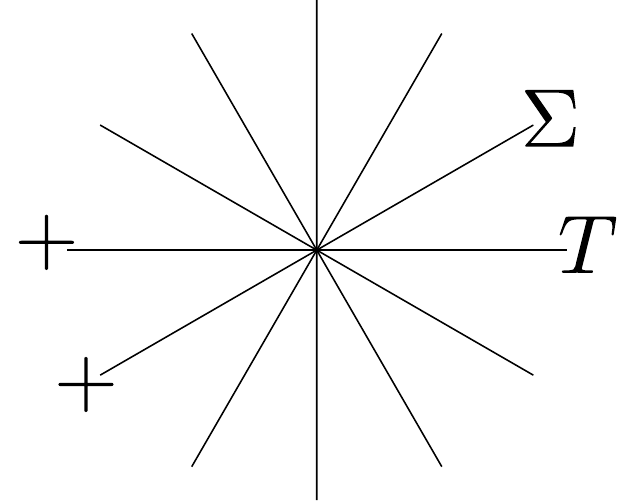}
        \caption{}
        \label{fig:parity_A2}
    \end{subfigure}

    \vspace{0.2cm}

    \begin{subfigure}{0.48\linewidth}
        \centering
        \includegraphics[width=\linewidth]{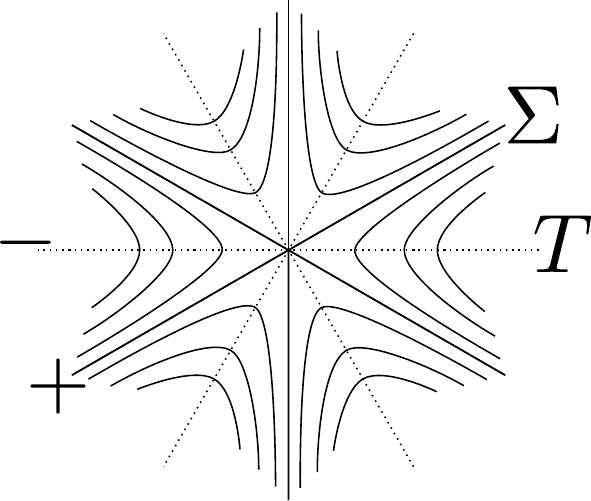}
        \caption{}
        \label{fig:parity_B1}
    \end{subfigure}
    \hfill
    \begin{subfigure}{0.48\linewidth}
        \centering
        \includegraphics[width=\linewidth]{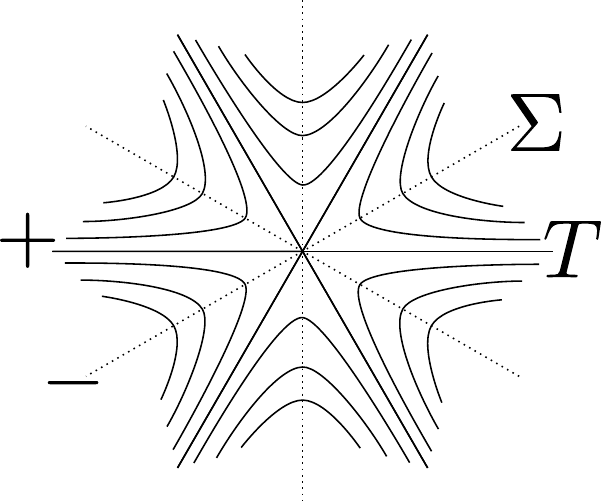}
        \caption{}
        \label{fig:parity_B2}
    \end{subfigure}
    \caption{The four possible methods by which the parity can decompose. Solid lines represent the far-field polarisation, while dotted lines represent the high symmetry lines with odd parity. (a) and (b) possess a \gls{tc} of 1, while (c) and (d) possess a \gls{tc} of -2.}
    \label{fig:parity_main}
\end{figure}

The compatibility relations, which determines how each $\Gamma$ point mode decomposes along the $\mathrm{T}$ and $\Sigma$ direction is shown in Table \ref{tab:c6v_compatibility} \cite{sakoda2005}. The $A$ ($B$) mode has even (odd) parity to the magnetic field, which results in an odd (even) parity to the far-field polarisation (as it transforms equivalently to the electric field). While Table \ref{tab:c6v_compatibility} only determines the parity along $T$ and $\Sigma$ direction, the rotational symmetry of the structure means that there are six equivalent high symmetry lines, and therefore the \gls{tc} can be determined from parity considerations up to a $6n$ ambiguity. The four possible decompositions for the fundamental ($n=0$) states are shown in Figure \ref{fig:parity_main}.

\begin{table}[ht] 
    \renewcommand{\arraystretch}{1.2}
    \begin{tabular}{c|cccccc} 
        \hline 
          & T & $\Sigma$ \\ 
        \hline 
        $A_1$ & $A \hspace{0.5em}(-1)$ & $A \hspace{0.5em}(-1)$ \\ 
        $A_2$ & $B \hspace{0.5em}(1)$ & $B \hspace{0.5em}(1)$ \\ 
        $B_1$ & $A \hspace{0.5em}(-1)$ & $B \hspace{0.5em}(1)$ \\ 
        $B_2$ & $B \hspace{0.5em}(1)$ & $A \hspace{0.5em}(-1)$ \\ 
        $E_1$ & $A+B \hspace{0.5em}(-1 + 1)$ & $A+B \hspace{0.5em}(-1 + 1)$ \\ 
        $E_2$ & $A+B \hspace{0.5em}(-1 + 1)$ & $A+B \hspace{0.5em}(-1 + 1)$ \\ \hline 
    \end{tabular} 
    \caption{Compatibility relations for the $\Gamma$ point in the $C_{6v}$ space group. The parity of the far-field polarisation for each mode is provided within the brackets, where $+1$ represents an even parity and $-1$ represents an odd parity.} 
    \label{tab:c6v_compatibility}
\end{table}

By combining the information from Table \ref{tab:c6v_compatibility} and Figure \ref{fig:parity_main}, we are able to state that the far-field polarisation of the $A_1$ mode is given by Figure \ref{fig:parity_A1}, while the far-field polarisation of the $A_2$ mode is shown in Figure \ref{fig:parity_A2}. Both modes correspond to a \gls{tc} of 1. Meanwhile, the far-field polarisation of the $B_1$ and $B_2$ modes, given by Figures \ref{fig:parity_B1} and \ref{fig:parity_B2}, both correspond to a \gls{tc} of -2. The \gls{tc} from this approach is consistent with considering the change in the eigenvalue of the structure under a $C_6$ rotation \cite{zhen2014, salerno2022}; meanwhile the far-field polarisation direction in reciprocal space is equivalent to the direction of the electric field in real space, as can be seen from comparing Figures \ref{fig:c6v_fields} and \ref{fig:parity_main}.

There are two different methods by which a doubly degenerate mode can decompose into an $A + B$ basis along the $T$ and $\Sigma$ directions. Firstly the degenerate mode can split into one mode that has odd parity along both high symmetry lines, and one mode that has even parity along both high symmetry lines \cite{weimin2024}. Both of these modes would have a \gls{tc} of 1, and would be of the form seen in Figure \ref{fig:parity_A1} and \ref{fig:parity_A2}. Secondly the degenerate mode could split into two modes where both modes have an even parity along one high symmetry line, and an odd parity along the other. This would lead to a far-field distribution as seen in Figures \ref{fig:parity_B1} and \ref{fig:parity_B2}, both of which possess a \gls{tc} of -2. The two possible values of the \gls{tc} are contrary to what was expected from the calculation of their eigenvalue under rotational symmetry \cite{salerno2022} -- this discrepancy is due to the undefined nature of the far-field polarisation at the $\Gamma$ point. 

By comparing to simulation results \cite{chern2026, ratiu2026} we can state that the $E_1$ mode is the \gls{irrep} with a \gls{tc} of 1, while the $E_2$ \gls{irrep} has a \gls{tc} of -2. We'd also expect the magnitude of the \gls{tc} of the $E_2$ \gls{irrep} to be twice that of the $E_1$ \gls{irrep} due to their respective angular momentum basis, which is consistent with the results found here. We may also explain the \gls{tc} of both modes using the $\bm{k} \cdot \bm{p}$ perturbation theory. 

\subsection{\texorpdfstring{$\bm{k}\cdot\bm{p}$}{k dot p} Perturbation Theory}
While the reflection parity method can be used to determine the \gls{tc} of each of the modes, it is unable to determine the scaling of the \gls{q} factor as the $\bm{k}$ vector is moved away from the $\Gamma$ point. For that, a perturbation theory approach is required. 

The eigenmodes within a \gls{phc} structure at $\bm{k}$ vectors near the $\Gamma$ point satisfy \cite{ochiai2024}:
\begin{equation}
    \mathcal{H} \ket{\psi} = \mathcal{E} \ket{\psi}, \qquad  \mathcal{E} = \frac{\omega^2}{c^2},
\end{equation}
where the Hamiltonian is a sum of the Hamiltonian at the $\Gamma$ point, and perturbation terms of progressively higher order of $\bm{k}$. The perturbation may be treated as a perturbation by an $E_1$ \gls{irrep}, as $(k_x, k_y)$ transforms equivalently to the basis function $(x, y)$ of the $E_1$ \gls{irrep}. The effective Hamiltonian can then be derived using the symmetry relation under point-group operations \cite{ochiai2024}:
\begin{equation}
    \mathcal{H}_{R_1 R_2} (\bm{k}) = D_{R_1}^\dagger(A) \mathcal{H}_{R_1 R_2}(A\bm{k})D_{R_2}(A),
    \label{eq:symmetry_relation}
\end{equation}
where $D_R(A)$ is the representation matrix of \gls{irrep} $R$ under symmetry operation $A$. Note that when looking at the perturbation to the $E_2$ mode the two representation matrices are representation matrices of different \glspl{irrep}, as using the same representation matrix will lead to an incorrect Hamiltonian \cite{long2023}. The Hamiltonian for the $E_1$ and $E_2$ modes were derived up to a third order term in Appendix \ref{app:e1_e2}, and were found from Equations \ref{eq:appendix_E1_hamiltonian} \cite{guo2020} and \ref{eq:appendix_E2_hamiltonian} to be:
\begin{equation}
    \mathcal{H}_{E_1} =  (\omega - i\gamma_0 + a \left|k\right|^2) \mathbf{I} + 
    b(k_x^2 - k_y^2) \sigma_z + 2bk_xk_y \sigma_x,
    \label{eq:E1_hamiltonian}
\end{equation}
\begin{multline}
    \mathcal{H}_{E_2} = (\omega - i\gamma_0 + d(k_x^3 - 3k_xk_y^2)) \mathbf{I} + {} \\
    (a + b(k_x^2 + k_y^2))(k_x \sigma_z - k_y\sigma_x) - c(k_y^3 - 3k_x^2k_y)\sigma_y,
    \label{eq:E2_hamiltonian}
\end{multline}   
where all terms are defined in Appendix \ref{app:e1_e2}. From Equation \ref{eq:E1_hamiltonian} we can see that the perturbation to the $E_1$ mode introduces terms with the symmetry of the $A_1$ and $E_2$ \glspl{irrep}. The result is consistent with the direct product of the $E_1$ mode with itself, seen in Table \ref{tab:c6v_products}, as the $A_2$ mode only exists for terms of sixth order or higher. 

\begin{table}[ht] 
    \renewcommand{\arraystretch}{1.2}
    \setlength{\tabcolsep}{4pt}
    \begin{tabular}{c|cccccc} 
        \hline $C_{6v}$ & $A_1$ & $A_2$ & $B_1$ & $B_2$ & $E_1$ & $E_2$ \\ 
        \hline $A_1$ & $A_1$  & $A_2$  & $B_1$ & $B_2$ & $E_1$ & $E_2$ \\ 
        $A_2$ & & $A_1$ & $B_2$ & $B_1$ & $E_1$ & $E_2$ \\ 
        $B_1$ &  &  & $A_1$ & $A_2$ & $E_2$  & $E_1$ \\ 
        $B_2$ &  &  &  & $A_1$ & $E_2$ & $E_1$ \\ 
        $E_1$ &  &  &  &  & $A_1 + A_2 + E_2$ & $B_1 + B_2 + E_1$ \\ 
        $E_2$ & & & & & & $A_1 + A_2 + E_2$ \\ \hline
    \end{tabular} 
    \caption{Direct product table of the $C_{6v}$ space group. Note that the symmetry arguments of the direct product is commutative.} 
    \label{tab:c6v_products}
\end{table}

Meanwhile, Equation \ref{eq:E2_hamiltonian} shows the effect of the perturbation on the $E_2$ mode. We can see that the first order terms possess the symmetry of the radiative $E_1$ mode, allowing coupling to the far-field radiation. Third order terms provide further coupling to the $E_1$ mode, as well as coupling to both the $B_1$ and $B_2$ mode. Note that the coupling constant to the $B_1$ and $B_2$ mode are different because of their non-degenerate nature, while a single coupling constant applies for the two-dimensional $E_1$ mode. These results are again consistent with the expectation from the direct products shown in Table \ref{tab:c6v_products}. 

We may use the symmetry of the $\bm{k} \cdot \bm{p}$ perturbation, which can also be determined through the direct product table, to determine how the \gls{q} factor of each mode scales as the magnitude of the $\bm{k}$ vector moves away from the $\Gamma$ point. For a mode to couple to the far-field it must have the same symmetry as the far-field radiation, that is the symmetry of the $E_1$ mode. As the $E_1$ mode contains that symmetry at the $\Gamma$ point it is already radiative and its \gls{q} factor is therefore generally unchanged as $\left|\bm{k}\right|$ is increased. Meanwhile, the first-order $\bm{k} \cdot \bm{p}$ perturbation of the $A_1, A_2 \text{ and } E_2$ modes introduces an $E_1$-symmetric term into the Hamiltonian \cite{ochiai2024}. This term causes partial symmetry-matching between the cavity mode and the far-field radiation, enabling coupling to the far-field radiative mode. Since the $E_1$-symmetric term appears at first order in $\bm{k}$, the coupling strength scales linearly with $\bm{k}$, resulting in a quality factor that scales as $Q \propto 1/k^2$ \cite{zhang2025}. Finally, the Hamiltonian of both $B$ modes do not contain an $E_1$-symmetric term under a first-order $\bm{k} \cdot \bm{p}$ perturbation, but instead an $E_2$-symmetric term is introduced, as shown in Equation \ref{eq:B_coupling_matrices}. To couple to the far-field a second $\bm{k} \cdot \bm{p}$ perturbation must be applied to transform the $E_2$-symmetric term into an $E_1$-symmetric term. Since the $E_1$-symmetric term therefore appears at second order in $\bm{k}$, the coupling strength scales quadratically with $\bm{k}$ resulting in a higher order $Q \propto 1/k^4$ dependence.

We can therefore state that, for two-dimensional \glspl{irrep} there is no direct relation between the \gls{tc} of the mode and the decay of the \gls{q} factor, and the $Q \propto 1/|k|^{2|q|}$ relation often quoted in the literature \cite{jin2019, kang2022} does not necessarily hold. Rather, the \gls{tc} of the mode must be determined through either its parity under reflection, or which modes it decomposes to under a perturbation. Meanwhile, the scaling of the \gls{q} factor should be determined by the order of $\bm{k}$ in the $\bm{k} \cdot \bm{p}$ perturbation required to reach the radiative mode. 

\section{\label{sec:numerical_simulations}Numerical Simulations}
\subsection{Method}
\begin{figure}[ht]
    \centering
    \includegraphics[width=0.8\linewidth]{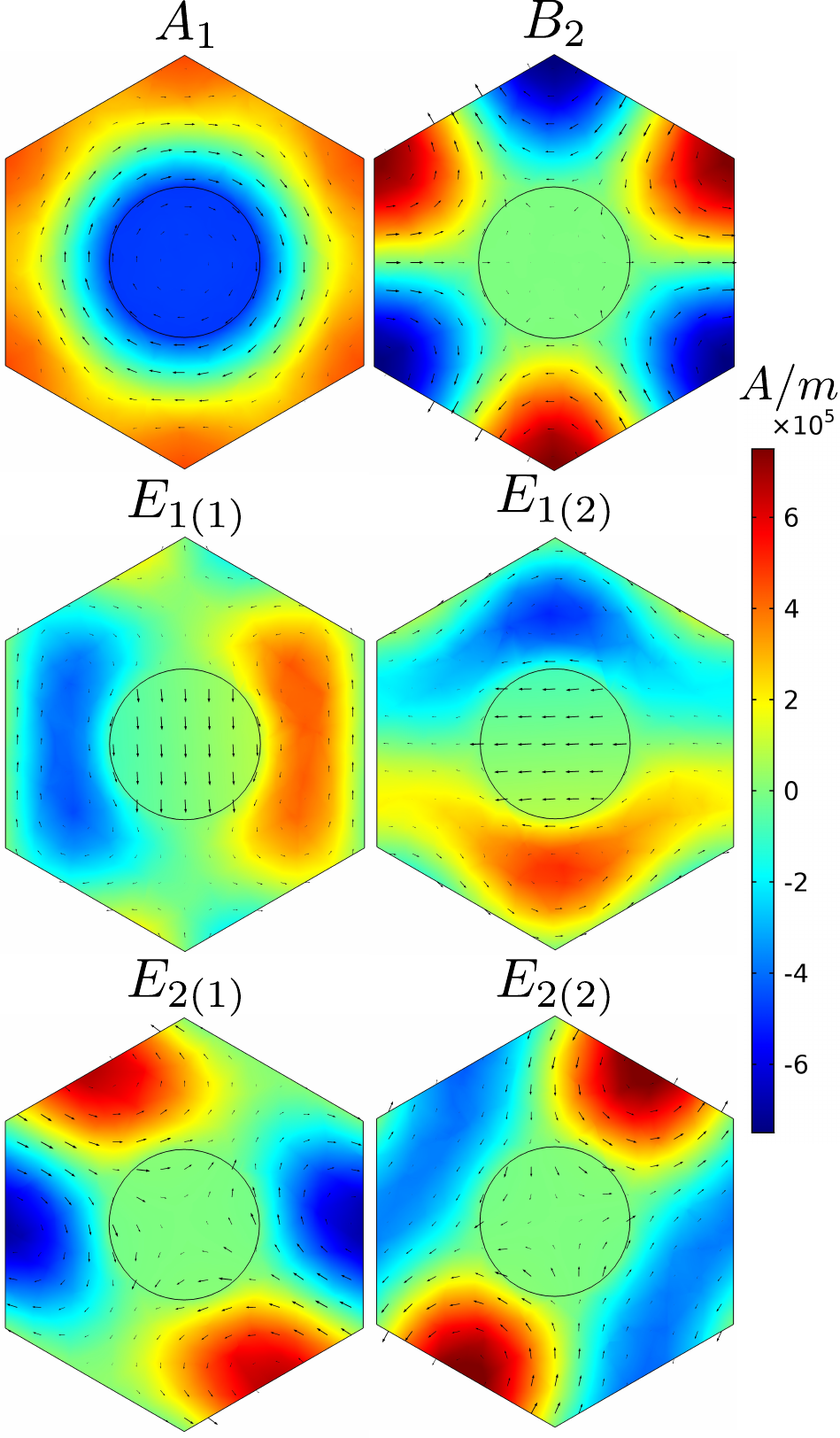}
    \caption{Near-field distribution of the unit cell in the six bound resonance / \gls{bic} modes at the $\Gamma$ point. The colour plot shows the distribution of the imaginary component of the magnetic field, while the arrows represent the direction of the electric field.}
    \label{fig:numerical_near_field}
\end{figure}

The theoretical results derived in the previous sections may be verified through the use of numerical simulations. Numerical simulations were conducted using the COMSOL Multiphysics \gls{fem} solver. A unit cell of the structure, seen in Figure \ref{fig:real_symmetries}, was simulated with Bloch periodic boundary conditions. The relative permittivity of the centre circle was set to 1, while the surrounding material was set to a relative permittivity of 11.7 (the permittivity of Si at \qty{1550}{\nm}). The height of the cavity was set to $0.4a$, while an air structure of height $5a$ was placed on top of the cavity. A \gls{pmc} boundary condition was set at the bottom of the cavity, which can be used to find the fundamental \gls{te} slab modes of the structure due to the $z$ inversion symmetry. A \gls{pml} was set on top of the simulation region to simulate a closed structure, before an eigenvalue solver was used to determine the eigenvalue and vectors of the structure. The band structure (Figure \ref{fig:band_structure}) was calculated from the eigenfrequencies of the moderate-to-high-Q solutions at the $\bm{k}$ vectors along the high symmetry lines (Figure \ref{fig:reciprocal_symmetries}). 

\begin{figure}[!t]
    \begin{subfigure}{0.9\linewidth}
        \includegraphics[width=\linewidth]{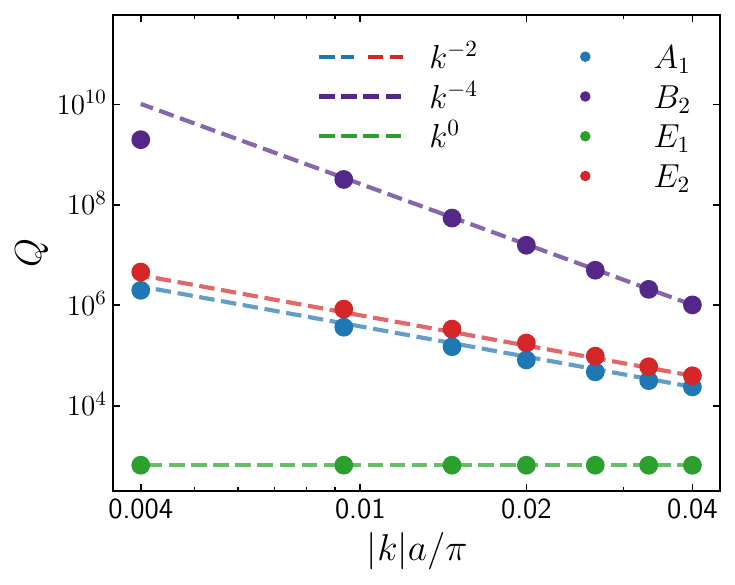}
        \caption{}
        \label{fig:log_log_plot}
    \end{subfigure}

    \begin{subfigure}{0.95\linewidth}
        \includegraphics[width=\linewidth]{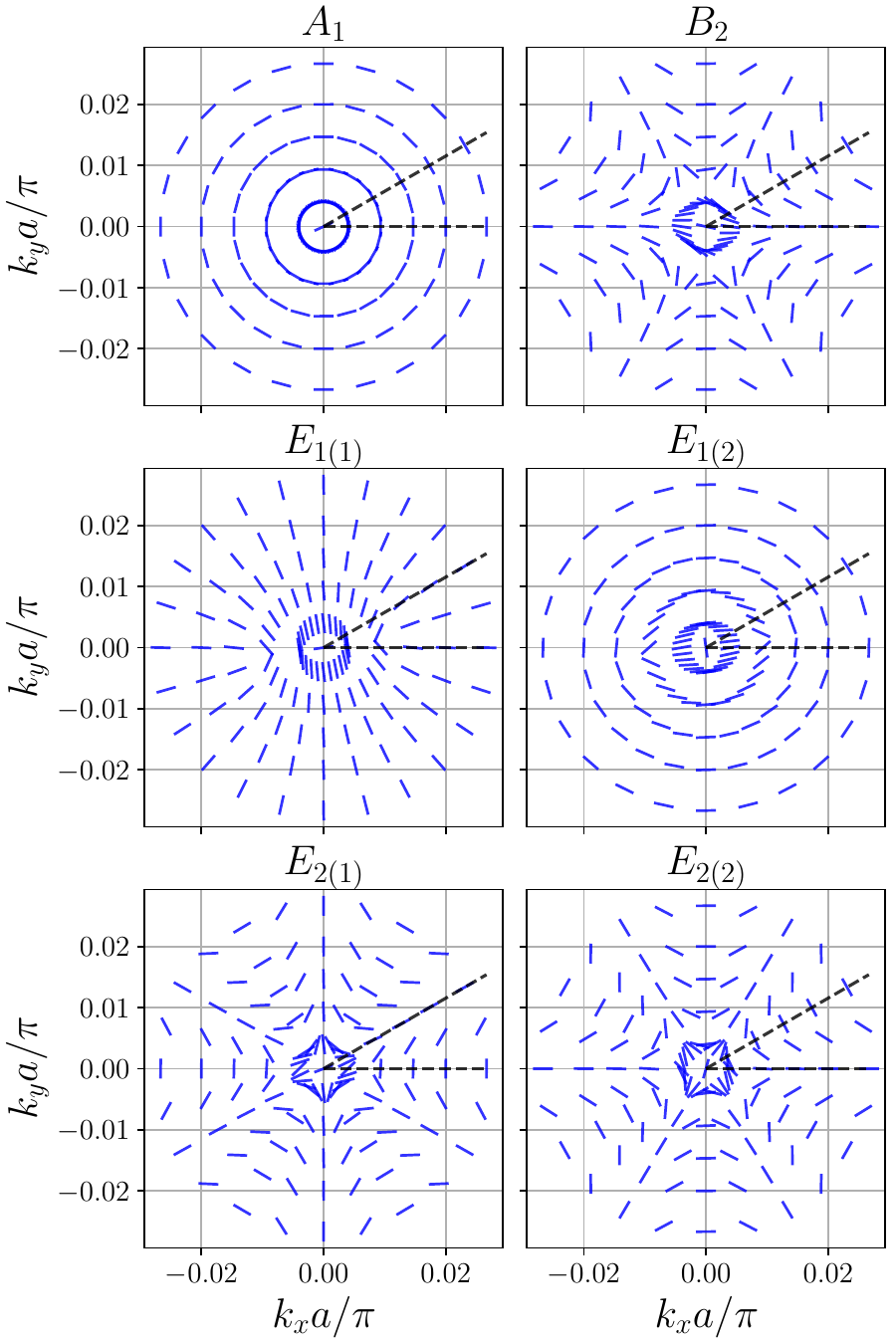}
        \caption{}
        \label{fig:numerical_far_field}
    \end{subfigure}
    \caption{Numerical results calculated at different $\bm{k}$ vectors from \glsentryshort{fem} simulations. (\ref{sub@fig:log_log_plot}) Scaling of the \gls{q} factor at increasing magnitudes of $\left|\bm{k}\right|$. (\ref{sub@fig:numerical_far_field}) Distribution of the far-field polarisation direction. Black dashed lines represent the high symmetry $\Sigma$ and $T$ lines at which the polarisation direction is either perpendicular or parallel.}
\end{figure}

The structure was simulated with the $\bm{k}$ vector for the Bloch boundary condition set to 0, to determine the near field distribution at the $\Gamma$ point. From the near field distribution it was possible to determine the \glspl{irrep} of each mode. Next, the $\bm{k}$ vector was changed in a ring of progressively increasing radius around the $\Gamma$ point to determine the far-field distribution of each of the \glspl{irrep}. 

The scaling of the \gls{q} factor as the $\bm{k}$ vector is moved away from the $\Gamma$ point can be calculated by writing the \gls{q} factor as:
\begin{equation}
Q = C|k|^\alpha,
\end{equation}
where $r_k$ is the radius of the $\bm{k}$ vector, $\alpha$ is the power with which \gls{q} scales, and $C$ is an arbitrary constant. $Q$ is determined numerically by $Q = \Real(\nu) / (2 \times \Imag(\nu))$. Taking the natural logarithm of each side, and re-writing in terms of the equation of a straight line, we find that the scaling of the \gls{q} factor at a particular radius in momentum space can be approximated as:
\begin{equation}
    \alpha = \dfrac{\ln(Q_2) - \ln(Q_1)}{\ln(k_2) - \ln(k_1)}.
\end{equation}

\subsection{Results}
By comparing Figure \ref{fig:c6v_fields} and \ref{fig:numerical_near_field} we are able to recognise the \gls{irrep} of each of the six zeroth order eigenfrequencies. We are also able to see that, except for slight differences due to the exact refractive index distribution, the field distribution for the $A_1$, $B_2$, and both $E_1$ modes are equivalent, which is due to the symmetry determining the basis function of the modes. While the $E_2$ modes are not immediately equivalent, this is due to the fact that any linear combination of the two basis functions are acceptable eigenvalues. Were an alternative linear combination chosen, we would see an equivalence between the theoretical and numerical results. 

The far-field distribution of each of the modes is plotted in Figure \ref{fig:numerical_far_field}. For all modes except $A_1$ the far-field distribution at the smallest $|\bm{k}|$ is undefined, for the $B_2$ mode this is because the far-field polarisation is still negligible, while for the other two modes it is because the perturbation is negligible and the modes are still degenerate. We will therefore focus on only the three largest $|\bm{k}|$ when discussing the far-field distribution. 

The far-field distribution of the non-degenerate modes are equivalent to the distribution of the electric field in the near-field (compare Figure \ref{fig:numerical_near_field} and \ref{fig:numerical_far_field}). We may therefore state that for non-degenerate modes it is possible to determine the \gls{tc} of the structure by only considering the near-field distribution. For the two degenerate modes the picture is more complicated. The $E_1$ modes decompose into two modes which have the same far-field distribution as the $A_1$ and $A_2$ \glspl{irrep}. These are the two non-degenerate \glspl{irrep} which are part of the direct product of the $E_1$ mode with the $\bm{k} \cdot \bm{p}$ perturbation (Table \ref{tab:c6v_products}). Meanwhile, the $E_2$ \gls{irrep} decomposes into two modes, one of which has the form of the $B_1$ mode, and one of which has the form of the $B_2$ mode. Both modes have a \gls{tc} of -2. The two modes are the modes contained within the direct product of the $E_2$ mode with the $\bm{k} \cdot \bm{p}$ perturbation. We may therefore state that the far-field distribution of the degenerate modes are also equivalent to a near field distribution, but the near field of a mode contained within the perturbation and not the near field of the original mode.

As the \gls{tc} of a structure is conserved under small perturbations it is possible to introduce a symmetry breaking perturbation which breaks the D point of charge -2 into a V point and numerous C points \cite{weimin2024}. Such a method could generate a \gls{bic} where adjacent $\bm{k}$ vectors emit circularly polarised light. However, the optimisation required for the generation of such a structure will be left as an avenue of future research. 

\begin{table}[ht] 
    \renewcommand{\arraystretch}{1.2}
    \setlength{\tabcolsep}{4pt} 
    \begin{tabular}{c|cccc} 
        \hline & Basis Vectors & \gls{tc} & Direct Product & Q Scaling  \\ 
        \hline $A_1$ & $x^2+y^2$ & 1 & {\color{blue}$E_1$} & $1/k^2$  \\ 
        $A_2$ & & 1 & {\color{blue}$E_1$} & $1/k^2$  \\ 
        $B_1$ & $x^3 -3xy^2$ & -2 & $E_2$ & $1/k^4$  \\ 
        $B_2$ & $y^3 - x^2y$ & -2 & $E_2$ & $1/k^4$  \\ 
        $E_1$ & $(x, y)$ & 1 & $\bm{A_1}+\bm{A_2}+E_2$ & 0  \\ 
        $E_2$ & $(x^2-y^2, -2xy)$ & -2 & $\bm{B_1}+\bm{B_2} + {\color{blue}E_1}$ & $1/k^2$  \\ 
        \hline 
    \end{tabular} 
    \caption{Summary of the information derived for each of the six \glspl{irrep} for a structure with $C_{6v}$ symmetry. Basis vectors are for the fundamental magnetic field distribution in real space. The bolded element in the direct product column represent the symmetry of the far-field polarisation, while the elements in blue represent the radiative coupling.} 
    \label{tab:summary} 
\end{table}

When determining the scaling of the \gls{q} factor we consider $\bm{k}$ vectors away from the accidental-\glspl{bic}. By looking at Figure \ref{fig:log_log_plot} we find that there is no relationship between the $\bm{k}$ vector and the \gls{q} factor for the $E_1$ mode as it is the radiative mode at the $\Gamma$ point. Note that for both the $E_1$ and $E_2$ modes the \gls{q} factor of the two degenerate modes are indistinguishable and only one of the two modes has been plotted. Meanwhile, the $A_1$ and $E_2$ modes scale as $Q \propto 1/|\bm{k}|^2$, while the $B_2$ mode scales as $Q \propto 1/|\bm{k}|^4$. All of these results are as expected from the $\bm{k} \cdot \bm{p}$ perturbation theory. Table \ref{tab:summary} summarises the information derived for each of the \glspl{irrep} of the $C_{6v}$ space group. 

\section{Conclusion}
\glsreset{tc}
The symmetry properties of a \gls{phc} with $C_{6v}$ symmetry has been thoroughly described. Near-field distribution of each \gls{irrep} of the space group can be described using an angular momentum basis, where the real and imaginary component of the basis functions are treated distinctively for the non-degenerate and doubly-degenerate cases. In the far-field, three different types of polarisation singularities, the V, C and D points have been introduced, and it was shown that a \gls{tc} in the far field does not necessarily indicate the presence of a \gls{bic}. All modes within the $C_{6v}$ structure were shown to be either V or D points, while introducing a symmetry-breaking perturbation leads to a C point due to the conserved nature of the \gls{tc}. 

The parity of the far-field polarisation along high symmetry lines was used to show that the \gls{tc} of the $A_1$, $A_2$ and $E_1$ modes is 1, while the $B_1$, $B_2$ and $E_2$ modes have a \gls{tc} of -2. Using the $\bm{k} \cdot \bm{p}$ perturbation theory the Hamiltonian of the $E_2$ mode was derived up to a third order perturbation for the first time. This showed that the perturbation of the $E_2$ mode had the symmetry of the $B_1 + B_2 + E_1$ basis, while the perturbation of the $E_1$ mode introduced the symmetry of the $A_1 + A_2 + E_2$ \glspl{irrep}; both Hamiltonians matched the symmetry expected from the direct product of \glspl{irrep}, and the symmetry of the perturbation directly determines the far-field polarisation of the degenerate modes. Perturbation theory was also used to show that the oft-quoted $Q \propto 1/|k|^{2|q|}$ dependence is untrue for the degenerate modes, and the perturbations that include the radiative mode should instead be used to determine the scaling of the \gls{q} factor. 

All theoretical results were then verified using the \gls{fem} simulation method for a hexagonal unit cell with a single air hole; the same symmetry arguments can hold for both the honeycomb and kagome lattices. Future research could include looking at symmetry-breaking perturbations for the formation of simultaneous C and V points, leading to the generation of a \gls{bic} which emits circularly polarised light at adjacent wavevectors. 

\subsection*{Acknowledgments}
{\small
This work was supported by the EPSRC funded Compound Semiconductor Manufacturing Hub for a Sustainable Future (Grant No. EP/Z532848/1) and the EPSRC CDT in Compound Semiconductor Manufacturing (EP/Y035801/1).}

\bibliography{references_bibtex}% Produces the bibliography via BibTeX.

\clearpage
\onecolumngrid
\appendix
\begin{center}
    \LARGE \bfseries Appendix
\end{center}
\addcontentsline{toc}{section}{Appendix}
\renewcommand{\theequation}{\thesection\arabic{equation}}
\setcounter{equation}{0}
\section{\label{app:e1_e2}Matrix Elements for \texorpdfstring{$E_1$ and $E_2$}{E1 and E2} Modes}
The Hamiltonian for the perturbation of the $E_1$ mode has been previously derived and is given by \cite{guo2020}:
\begin{equation}
    \mathcal{H} = 
    \begin{pmatrix}
        \omega_0 - i \gamma_0 + a |\bm{k}|^2 + b(k_x^2-k_y^2) & 2bk_x k_y \\
        2 b k_x k_y & \omega_0 - i \gamma _0 + a |\bm{k}|^2 - b (k_x^2 - k_y^2)
    \end{pmatrix},
    \label{eq:appendix_E1_hamiltonian}
\end{equation}
where $a=A-iA', b=B-iB'$ are complex coefficients, and the real and imaginary parts of the Hamiltonian are Hermitian. $\omega_0$ and $\gamma_0$ are the resonant wavelength and loss of the system at the $\Gamma$ point. Both the $A_1$ and $E_2$ modes are clearly present in the Hamiltonian through the $|\bm{k}|^2$ and $(k_x^2-k_y^2, 2k_x k_y)$ terms (the $A_2$ mode only appears at higher orders). While the derivation in the referenced article is provided only up to second order, it can be seen from equation S6 in \cite{guo2020} that the third order terms must all be 0.

While some efforts have been made to derive the equivalent Hamiltonian for the $E_2$ mode, these have either wrongly defined the mode of the $\bm{k} \cdot \bm{p}$ perturbation \cite{long2023}, or have only looked at the perturbation up to first order \cite{ochiai2024}. A full derivation of the Hamiltonian will therefore be provided here. 

The symmetry relation under point-group operations is crucial for the derivation, and is given by \cite{ochiai2024}:
\begin{equation}
    \mathcal{H}_{R_1 R_2} (\bm{k}) = D_{R_1}^\dagger(A) \mathcal{H}_{R_1 R_2}(A\bm{k})D_{R_2}(A),
    \label{eq:symmetry_relation_appendix}
\end{equation}
where $R_1$ and $R_2$ are the two \glspl{irrep} that are being coupled, while $D_R(A)$ is the representation matrix of the \gls{irrep} $R$ under symmetry operation $A$. It is possible to re-write Equation \ref{eq:symmetry_relation_appendix} for the perturbation of the $E_2$ mode by the $\bm{k}$ vector that transforms as the $E_1$ \gls{irrep} as:
\begin{equation}
    D_{E_1}(A) \mathcal{H}_{E_1 E_2}(\bm{k})D_{E_2}^\dagger(A) = \mathcal{H}_{E_1 E_2}(A\bm{k}).
\end{equation}
The general form of the Hamiltonian must follow \cite{long2023}:
\begin{equation}
    \mathcal{H}(\bm{k}) = (\omega_0 - i\gamma_0)\mathbf{I} + A(\bm{k}) - i B(\bm{k}),
\end{equation}
where both $A$ and $B$ are $2\times2$ Hermitian matrices. As both matrices will have the same general symmetry form, we can focus only on the symmetry of the $A$ matrix. As a Hermitian matrix, we may write it without loss of generality as:
\begin{equation}
    A(\bm{k}) = 
    \begin{pmatrix}
        h(\bm{k}) + g(\bm{k}) & f(\bm{k}) \\
        f^*(\bm{k}) & h(\bm{k}) - g(\bm{k})
    \end{pmatrix}.
\end{equation}
For a perturbation up to third order $h(\bm{k})$ can be written as:
\begin{equation}
    h(k_x, k_y) = h_{10} k_x + h_{01} k_y + h_{20} k_x^2 + h_{11} k_x k_y + h_{02} k_y^2 + h_{30} k_x^3 + h_{21} k_x^2 k_y + h_{12} k_x k_y^2 + h_{03} k_y^3
    \label{eq:general_h}
\end{equation}
while equivalent expressions exist for $g(\bm{k})$ and $f(\bm{k})$. 

We can determine the form of each element of the Hamiltonian by determining their transformation under each symmetry operation. The basis vectors of each of the two \glspl{irrep} are defined as:

\begin{equation} 
    \ket{E_1} = 
    \begin{pmatrix} 
        k_x \\ k_y 
    \end{pmatrix} 
    \qquad 
    \ket{E_2} = 
    \begin{pmatrix} 
        k_x^2 - k_y^2 \\ -2 k_x k_y 
    \end{pmatrix}.
    \label{eq:basis_vectors} 
\end{equation}

\subsection{Two-fold Rotation Symmetry \texorpdfstring{$C_2$}{C2}}
The two representation matrices for the $C_2$ rotation are:
\begin{equation}
    D_{E_1}(C_2) = 
    \begin{pmatrix}
        -1 & 0 \\ 0 & -1
    \end{pmatrix},
    \qquad
    D_{E_2}(C_2) = 
    \begin{pmatrix}
        1 & 0 \\ 0 & 1
    \end{pmatrix}
    = D_{E_2}^\dagger(C_2).
\end{equation}
Applying the expression to Equation \ref{eq:symmetry_relation_appendix} produces the following expression:
\begin{multline} 
    - \begin{pmatrix} 
        h(k_x, k_y) + g(k_x, k_y) & f(k_x, k_y) \\ 
        f^*(k_x, k_y) & h(k_x, k_y) - g(k_x, k_y) 
    \end{pmatrix} 
    = {} \\
    \begin{pmatrix} 
        h(-k_x, -k_y) + g(-k_x, -k_y) & f(-k_x, -k_y) \\ 
        f^*(-k_x, -k_y) & h(-k_x, -k_y) - g(-k_x, -k_y) 
    \end{pmatrix}.
\end{multline}
From this we can see that the constraint $h_{20} = h_{02} = h_{11} = 0$ applies, while an equivalent constraint holds for $f$ and $g$. No second-order terms may exist for the perturbation. 

\subsection{Reflection Symmetry Around the Vertical Plane \texorpdfstring{$\sigma_x$}{sigma x}}
The representation matrices for the $\sigma_x$ symmetry operation are:
\begin{equation}
    D_{E_1}(\sigma_x) = 
    \begin{pmatrix}
        -1 & 0 \\ 0 & 1
    \end{pmatrix},
    \qquad
    D_{E_2}(\sigma_x) =
    \begin{pmatrix}
        1 & 0 \\ 0 & -1
    \end{pmatrix}
    = D_{E_2}^\dagger(\sigma_x),
\end{equation}
which when applied to Equation \ref{eq:symmetry_relation_appendix} leads to:
\begin{multline}
    \begin{pmatrix}
        -\left[h(k_x, k_y) + g(k_x, k_y)\right] & f(k_x, k_y) \\
        f^*(k_x, k_y) & -\left[h(k_x, k_y) - g(k_x, k_y)\right]
    \end{pmatrix}
    = {} \\
    \begin{pmatrix}
        h(-k_x, k_y) + g(-k_x, k_y) & f(-k_x, k_y) \\
        f^*(-k_x, k_y) & h(-k_x, k_y) - g(-k_x, k_y)
    \end{pmatrix}.
\end{multline}
This provides further constraints on the matrix elements of $h_{01} = h_{21} = h_{03} = 0$, $f_{10} = f_{30} = f_{12} = 0$, with $g$ having the same constraints as $h$. Note that the symmetry operation $\sigma_y$ leads to the same constraints. 

We are therefore able to state that the form of the elements of the Hamiltonian have been reduced to:
\begin{subequations} 
    \begin{align} 
        h(k_x, k_y) &= h_{10} k_x + h_{30} k_x^3 + h_{12} k_x k_y^2, \\ 
        g(k_x, k_y) &= g_{10} k_x + g_{30} k_x^3 + g_{12} k_x k_y^2, \\ 
        f(k_x, k_y) &= f_{01} k_y + f_{21} k_x^2 k_y + f_{03} k_y^3. 
    \end{align} 
    \label{eq:elements_before_c6}
\end{subequations}

\subsection{Six-fold Rotation Symmetry \texorpdfstring{$C_6$}{C6}}
For the principal rotational symmetry, the $C_6$ rotation, we choose to consider a counter-clockwise rotation around the $\Gamma$ point, therefore the $\bm{k}$ vectors transform as:
\begin{equation}
    \begin{pmatrix}
        k_x' \\ k_y'
    \end{pmatrix}
    = 
    \begin{pmatrix}
        \frac{1}{2} & \frac{\sqrt{3}}{2} \\ -\frac{\sqrt{3}}{2} & \frac{1}{2}
    \end{pmatrix}
    \begin{pmatrix}
        k_x \\ k_y
    \end{pmatrix}.
\end{equation}
Meanwhile, due to the definition of the basis vectors in Equation \ref{eq:basis_vectors}, the representation matrices for the $C_6$ rotation are:
\begin{equation}
    D_{E_1}(C_6) = 
    \begin{pmatrix}
        \frac{1}{2} & \frac{\sqrt{3}}{2} \\ -\frac{\sqrt{3}}{2} & \frac{1}{2}
    \end{pmatrix},
    \qquad
    D_{E_2}(C_6) = 
    \begin{pmatrix}
        -\frac{1}{2} & -\frac{\sqrt{3}}{2} \\ \frac{\sqrt{3}}{2} & -\frac{1}{2}
    \end{pmatrix},
    \qquad
    D_{E_2}^\dagger(C_6) =
    \begin{pmatrix}
        -\frac{1}{2} & \frac{\sqrt{3}}{2} \\ -\frac{\sqrt{3}}{2} & -\frac{1}{2}
    \end{pmatrix}.
\end{equation}
Applying these equations to Equation \ref{eq:symmetry_relation_appendix} leads to the following:
\begin{multline}
    \begin{pmatrix}
        -\frac{\sqrt{3}}{2} \Real[f(k_x, k_y)] + \frac{g(k_x, k_y)}{2} - h(k_x, k_y) &
        \frac{f(k_x, k_y)}{2} - \frac{3i}{2} \Imag[f(k_x, k_y)] + \frac{\sqrt{3}}{2} g(k_x, k_y) \\
        \frac{f(k_x, k_y)}{2} + \frac{i}{2} \Imag[f(k_x, k_y)] + \frac{\sqrt{3}}{2} g(k_x, k_y) &
        \frac{\sqrt{3}}{2} \Real[f(k_x, k_y)] - \frac{g(k_x, k_y)}{2} - h(k_x, k_y)
    \end{pmatrix}
    = {} \\
    \begin{pmatrix}
        h(k_x', k_y') + g(k_x', k_y') & f(k_x', k_y') \\
        f^*(k_x', k_y') & h(k_x', k_y') - g(k_x', k_y')
    \end{pmatrix}.
\end{multline}
Starting with the linear terms we find for the top right element that:
\begin{equation} 
    \begin{aligned} 
        \frac{f_{01}}{2} k_y -& \frac{3i}{2} \Imag[f_{01} k_y] + \frac{\sqrt{3}}{2} g_{10} k_x = f_{01} \left( -\frac{\sqrt{3}}{2} k_x + \frac{k_y}{2} \right)
        \\[0.5em] 
        &\Rightarrow\qquad \Imag[f_{01}] = 0, \quad g_{10} = -f_{01}
    \end{aligned} 
    \label{eq:f01g10}
\end{equation}
where the two equations in the second line are derived by considering the $k_x$ and $k_y$ components separately. 

Proceeding to the term in the top left element we find:
\begin{equation}
    \begin{aligned}
        -\frac{\sqrt{3}}{2} f_{01} k_y &+ \frac{g_{10}}{2} k_x - h_{10} k_x = \left(h_{10} + g_{10} \right)  \left( \frac{k_x}{2} + \frac{\sqrt{3}}{2} k_y \right)
        \\ 
        &\Rightarrow\qquad \frac{g_{10}}{2} - h_{10} = \frac{h_{10}}{2} + \frac{g_{10}}{2} 
        \qquad - \frac{\sqrt{3}}{2} f_{01} = \frac{\sqrt{3}}{2} h_{10} + \frac{\sqrt{3}}{2} g_{10}
        \\
        &\qquad \Rightarrow  \qquad h_{10} = 0, \quad g_{10} = -f_{01}. 
    \end{aligned}
    \label{eq:h10}
\end{equation}

The same procedure can be applied to the third order perturbation. Starting again with the top right element we find:
\begin{multline}
    \frac{1}{2} \left(f_{21} k_x^2 k_y + f_{03} k_y^3 \right) - \frac{3i}{2} \Imag\left[f_{21} k_x^2 k_y + f_{03} k_y^3 \right] + \frac{\sqrt{3}}{2} \left(g_{30} k_x^3 + g_{12} k_x k_y^2 \right) = {} \\
    f_{21} \left( \frac{k_x}{2} + \frac{\sqrt{3}}{2} k_y \right)^2 \left( -\frac{\sqrt{3}}{2} k_x + \frac{k_y}{2} \right) + f_{03} \left( -\frac{\sqrt{3}}{2} k_x + \frac{k_y}{2} \right)^3 = {} \\
    \left(-\frac{\sqrt{3}}{8} f_{21} - \frac{3\sqrt{3}}{8} f_{03} \right) k_x^3 + \left( -\frac{5}{8} f_{21} + \frac{9}{8} f_{03} \right) k_x^2 k_y + {} \\
    \left( -\frac{\sqrt{3}}{8} f_{21} - \frac{3\sqrt{3}}{8} f_{03} \right) k_x k_y^2 + \left( \frac{3}{8}f_{21} + \frac{1}{8} f_{03} \right) k_y^3.
\end{multline}
From the $k_x^3$ and $k_x^2 k_y$ terms we find:
\begin{equation}
    \begin{aligned}
        \frac{\sqrt{3}}{2} g_{30} &= -\frac{\sqrt{3}}{8} f_{21} - \frac{3\sqrt{3}}{8} f_{03} \qquad
        \frac{f_{21}}{2} - \frac{3i}{2} \Imag[f_{21}] = -\frac{5}{8} f_{21} + \frac{9}{8} f_{03} \\
        &\Rightarrow\qquad \Imag[f_{21}] = - 3 \Imag[f_{03}], \quad \Real[f_{21}] = \Real[f_{03}], \quad g_{30} = -\Real[f_{03}].
    \end{aligned}
    \label{eq:f21f03g30}
\end{equation}
Additionally, looking at the constraint from the $k_x k_y^2$ term we find:
\begin{equation}
    \begin{aligned}
        \frac{\sqrt{3}}{2} g_{12} &= -\frac{\sqrt{3}}{8} f_{21} - \frac{3\sqrt{3}}{8} f_{03} \\
        &\Rightarrow\qquad g_{12} = -\Real[f_{03}] = g_{30}.
    \end{aligned}
    \label{eq:g12}
\end{equation}
No further constraints are found from the $k_y^3$ term. 

We may then look at the top left element and find:
\begin{multline}
    -\frac{\sqrt{3}}{2} \Real[f_{21} k_x^2 k_y + f_{03} k_y^3] + \frac{g_{30}}{2} \left(k_x^3 + k_x k_y^2 \right) - h_{30} k_x^3 - h_{12} k_x k_y^2 = {} \\
    h_{30} \left( \frac{k_x}{2} + \frac{\sqrt{3}}{2} k_y \right)^3 + h_{12} \left( \frac{k_x}{2} + \frac{\sqrt{3}}{2} k_y \right) \left( -\frac{\sqrt{3}}{2} k_x + \frac{k_y}{2} \right)^2 + {} \\
    g_{30} \left[ \left( \frac{k_x}{2} + \frac{\sqrt{3}}{2} k_y \right)^3 + \left( \frac{k_x}{2} + \frac{\sqrt{3}}{2} k_y \right) \left( -\frac{\sqrt{3}}{2} k_x + \frac{k_y}{2} \right)^2 \right] = {} \\
    \left( \frac{h_{30}}{8} + \frac{3}{8} h_{12} + \frac{g_{30}}{2} \right) k_x^3 + \left( \frac{3\sqrt{3}}{8} h_{30} + \frac{\sqrt{3}}{8} h_{12} + \frac{\sqrt{3}}{2} g_{30} \right) k_x^2 k_y + {} \\
    \left( \frac{9}{8} h_{30} - \frac{5}{8} h_{12} + \frac{g_{30}}{2} \right) k_x k_y^2 + \left( \frac{3\sqrt{3}}{8} h_{30} + \frac{\sqrt{3}}{8} h_{12} + \frac{\sqrt{3}}{2} g_{30} \right) k_y^3.
\end{multline}
Looking at the $k_x^3$ term we find the constraint:
\begin{equation}
    \begin{aligned}
        \frac{g_{30}}{2} -& h_{30} = \frac{h_{30}}{8} + \frac{3}{8} h_{12} + \frac{g_{30}}{2} \\
        & \Rightarrow\qquad h_{12} = -3 h_{30}.
    \end{aligned}
    \label{eq:h12h30}
\end{equation}
No additional constraints are found by considering the other $k$ terms. 

Combining Equation \ref{eq:elements_before_c6}, \ref{eq:f01g10}, \ref{eq:f21f03g30}, \ref{eq:g12}, \ref{eq:h10}, and \ref{eq:h12h30} the form of the matrix is found to be:
\begin{subequations}
    \begin{align}
        &h(k_x, k_y) = \delta k_x^3 - 3 \delta k_x k_y^2, \\
        &g(k_x, k_y) = \alpha k_x + \beta (k_x^3 + k_x k_y^2) = (\alpha + \beta \left|k\right|^2) k_x, \\
        &f(k_x, k_y) = -\alpha k_y - \beta (k_x^2 k_y + k_y^3) + i \gamma (k_y^3 - 3 k_x^2 k_y) = -(\alpha + \beta \left|k\right|^2) k_y + i \gamma (k_y^3 - 3 k_x^2 k_y), \\
        &A(\bm{k}) = (\alpha + \beta \left|k\right|^2) (k_x \sigma_z - k_y \sigma_x) + \delta (k_x^3 - 3 k_x k_y^2) \mathbf{I} - \gamma (k_y^3 - 3 k_x^2 k_y) \sigma_y.
    \end{align}
\end{subequations}

While these results are different from that derived in \cite{ochiai2024}, this can be explained by a different choice of $E_2$ basis function. From this we can see that the direct product of the $E_1$ and $E_2$ \glspl{irrep} leads to a Hamiltonian with a term representing the $E_1$ \gls{irrep} (the term multiplied by the coefficient $\alpha + \beta \left|k\right|^2$), one representing the $B_1$ \gls{irrep} ($\delta$ term), and one representing the $B_2$ \gls{irrep} ($\gamma$ term), as expected from the group theory arguments \cite{atkins1970}. The values of the coefficients $\alpha, \beta, \gamma, \delta$ can be determined from numerical simulation. As the $B$ matrix must follow the same symmetries we may write the full Hamiltonian of the system as:
\begin{equation}
    \mathcal{H} = 
    \begin{pmatrix}
        \omega - i\gamma_0 + (a + b\left|\bm{k}\right|^2)k_x + d(k_x^3 - 3k_xk_y^2) & -(a + b\left|\bm{k}\right|^2)k_y + i c(k_y^3 - 3k_x^2 k_y) \\
        -(a + b\left|\bm{k}\right|^2)k_y - i c(k_y^3 - 3k_x^2 k_y) & \omega - i\gamma_0 - (a + b\left|\bm{k}\right|^2)k_x + d(k_x^3 - 3k_xk_y^2)
    \end{pmatrix},
    \label{eq:appendix_E2_hamiltonian}
\end{equation}
where $a = \alpha - i \alpha', b = \beta - i \beta', c = \gamma - i \gamma', d = \delta - i \delta'$ are complex coefficients.

The eigenvalues of the system are then:
\begin{equation}
    \omega_\pm = \omega_0 - i \gamma_0 + d(k_x^3 - 3k_xk_y^2) \pm \sqrt{a^2 \left|\bm{k}\right|^2 + 2ab\left|\bm{k}\right|^4 + b^2\left|\bm{k}\right|^6 + c^2\left(k_y^3 - 3k_x^2 k_y\right)^2}.
\end{equation}

\section{Matrix Elements for \texorpdfstring{$B_1$ and $B_2$}{B1 and B2} Modes}
\setcounter{equation}{0}
We can use a similar approach for the B modes, however as one-dimensional modes the coupling matrix between the $E_1$ perturbation and the $B$ modes will not have a $2\times 2$ form which is convenient to write as a Hamiltonian. Instead, we may look at the coupling in the form of a $2 \times 1$ coupling matrix \cite{ochiai2024}:
\begin{equation}
    \mathcal{H}_{E_1 B_i} (\bm{k}) =
    \begin{pmatrix}
        h(\bm{k}) \\ g(\bm{k})
    \end{pmatrix}
    , \qquad i = 1, 2,
\end{equation}
where $h$ and $g$ have the same form as in Equation \ref{eq:general_h}. The representation matrix for the $B$ modes will be a single number, corresponding to the character of the mode under a particular rotation or reflection rather than a $2 \times 2$ matrix. Applying the same method of going through each symmetry transformation in turn results in the following coupling matrices for the $B_1$ and $B_2$ modes:
\begin{equation}
    \mathcal{H}_{E_1 B_1} (\bm{k}) =
    a
    \begin{pmatrix}
        k_x^2 - k_y^2 \\ -2 k_x k_y
    \end{pmatrix}
    , \qquad
    \mathcal{H}_{E_1 B_2} (\bm{k}) =
    b
    \begin{pmatrix}
        2 k_x k_y \\ k_x^2 - k_y^2
    \end{pmatrix}.
    \label{eq:B_coupling_matrices}
\end{equation}
We see that both matrices have the form of the $E_2$ \gls{irrep}, as expected from Table \ref{tab:c6v_products}. 

\end{document}

%% file: glossary_entries.tex
\newacronym{fsr}{FSR}{free spectral range}
\newacronym{pcsel}{PCSEL}{photonic crystal surface-emitting laser}
\newacronym{lidar}{LiDAR}{light detection and ranging}
\newacronym{fp}{FP}{Fabry–Pérot}
\newacronym{vcsel}{VCSEL}{vertical cavity surface-emitting laser}
\newacronym{phc}{PhC}{photonic crystal}
\newacronym{cwt}{CWT}{coupled wave theory}
\newacronym{ml}{ML}{machine learning}
\newacronym{fdtd}{FDTD}{finite-difference time-domain}
\newacronym{pwem}{PWEM}{plane-wave expansion method}
\newacronym{te}{TE}{transverse electric}
\newacronym{iqhe}{IQHE}{integer quantum Hall effect}
\newacronym{2deg}{2DEG}{two-dimensional electron gas}
\newacronym{pbg}{PBG}{photonic band gap}
\newacronym[longplural={bound states in the continuum},shortplural={BICs}]{bic}{BIC}{bound state in the continuum}
\newacronym[longplural={quasi-bound states in the continuum},shortplural={qBICs}]{qbic}{qBIC}{quasi-bound state in the continuum}
\newacronym{fw}{FW}{Friedrich-Wintgen}
\newacronym{bz}{BZ}{Brillouin zone}
\newacronym{q}{Q}{quality}
\newacronym{gr}{GR}{guided resonance}
\newacronym{tm}{TM}{transverse magnetic}
\newacronym{gme}{GME}{guided-mode expansion}
\newacronym{fem}{FEM}{finite-element method}
\newacronym{rsi}{RSI}{repetitive strain injury}
\newacronym{gt}{GT}{group theory}
\newacronym{pec}{PEC}{perfect electric conductor}
\newacronym{pmc}{PMC}{perfect magnetic conductor}
\newacronym{tc}{TC}{topological charge}
\newacronym{irrep}{irrep}{irreducible representation}
\newacronym{pml}{PML}{perfectly matched layer}
\newacronym{cp}{CP}{circularly polarised}
\newacronym{lcp}{LCP}{left circularly polarised}
\newacronym{rcp}{RCP}{right circularly polarised}